\documentstyle[prd,aps,epsfig,floats]{revtex}
\tighten
\begin{document}
\draft

\title
{Electron Screening in the $^7$Be $+$ p $\longrightarrow$ $^8$B $+$ $\gamma$ 
reaction}

\author{V.~B.~Belyaev${}^{1,2}$ \and D.~E.~Monakhov${}^1$ \and
D.~V.~Naumov${}^{1,3}$  and F.~M.~Pen'kov${}^1$}
\address{${}^1$ Joint Institute for Nuclear Research, Dubna, 141980, Russia \\
         ${}^2$ Research Center for Nuclear Physics, Osaka University,
         Japan \\
         ${}^3$ Physics Department, Irkutsk State University, Irkutsk,
         664003, Russia \\
         }
\date{\today}
\maketitle

\begin{abstract}
We evaluate the effect of screening by bound electron in 
${^7}$Be(p,$\gamma$)$^8$B reaction, where $^7$Be target contains bound
electron, in the framework of the adiabatic representation of 
the three particle problem. A comparison with two other approximations
(united atom and folding) is presented. A good agreement between the ``united
atom'' approximation and the exact solution is found. We also discuss the
screening corrections induced by two K-shell electrons on a $^7$Be target.
The bound electron screening effect consequences for $^7$Be and $^8$B solar
neutrino fluxes are discussed.
\end{abstract}

\section{Introduction}

In recent years, an increasing attention has been devoted to an accurate
estimation of electron screening effect for nuclear fusion reactions in
stellar plasma and for the interactions of low-energy ion 
beams with atomic or molecular targets in laboratory experiments (see
refs.~\cite{Gruzinov,Shoppa,Salpeter,Mitler,Carraro,Brown,Langanke,Shaviv} 
and references therein). 

In this Letter we present the first quantum mechanical calculation of
screening effect by bound electron in 
\begin{equation}
\label{creation}
{}^7\mathrm{Be}+\mathrm{p} \longrightarrow {}^8\mathrm{B}+\gamma
\end{equation}
nuclear fusion from the pp-cycle in the Sun. 
Contribution of this reaction into the the total luminosity of the
Sun is negligible small, but it is directly related to the long-standing
``Solar Neutrino Problem'', --  one of the most intriguing issue in the
present-day neutrino astrophysics.
Standard physics cannot explain an $^{37}$Ar production rate in the Chlorine
experiment smaller than that expected from the solar $^8$B neutrino flux
measured by both Kamiokande and (with better statistic) Super Kamiokande. 
GALLEX and SAGE experiments also indicate beryllium neutrino deficit 
(see the discussion in ref.~\cite{Innocenti}).  
One of the most elegant solutions to the solar neutrino anomaly is resonant 
neutrino flavor conversion in the sun, that is the so-called MSW
effect~\cite{MSW}. It requires an extension of the minimal standard
electroweak theory: neutrino masses and neutrino mixing. 
These neutrino oscillation parameters are determined in a way that can bridge 
between the predictions of the standard solar models and the solar neutrino 
observations.  Thus, even in the framework of the standard solar model within
a hypothesis of neutrino oscillation (and MSW effect), it is apparently needed
more precise calculations of nuclear fusion rates in the sun, because they 
can significantly affect the neutrino oscillation parameters determination.

A careful study of electron screening effect on nuclear fusion rates becomes
particularly actual in view of expected high accuracy neutrino flux 
measurements by a number of new large detectors (Super Kamiokande, SNO).
The interpretation of forthcoming data  requires relevant precise calculations
of  solar neutrino fluxes and neutrino energy spectrum.

Usually, the effects of surrounding plasma on the nuclear fusion are treated
in electrostatic screening approximation. This approximation, being 
classical or quantum, correctly reflects the major properties
of a process only for high relative velocity of the colliding nuclei, 
when electron density in the vicinity of the fusing nuclei remains almost
unchanged during the collision.
In the case, when relative velocity of the nuclei is much smaller or
comparable with the electron one (and this is the case at solar conditions), 
the electron density changes following any relative configuration of 
the nuclei, and the electrons have an impact on a kinetic energy shift 
of the nuclei. 
It is therefore natural to consider the phenomenon within the framework
of the adiabatic approximation, which comes from the well known
Born-Oppenheimer (BO) approximation.

The BO approximation allows one to treat nuclear motion independently from the 
electron coordinate, within a new effective potential which depends on the 
internuclear distance. Since, the considered nuclear velocities are smaller than
the boun electron one, corrections to the BO approximation are expected to be
negligible. Obviously, the fusing nuclei are from the continuum
energy spectrum. An accurate treatment within the adiabatic approximation of
the screening effect by electrons from continuum spectrum requires an 
additional research but the case of bound electrons presents no special
problem (see, for example ref.~\cite{Melezhik}).

As it was argued by A.~Dar, G.~Shaviv, and N.~Shaviv~\cite{Shaviv,Dar},
the commonly accepted Debye-H\"uckel theory is not quite adequate
for evaluating the screening effect in not-very-dense stars, like the Sun. 
There is, also, an experimental evidence that this theory does not provide 
correct answer for the screening~\cite{Shoppa}.
Actually, this fact is of no importance when the screening due to the plasma 
electrons is by itself rather small.
But it is not the case for the low-lying bound electrons which do screen the 
electric charge of nuclei much effectively, and moreover, the screening
effect drastically increases when energy of the fusing ions decreases.
The electron screening can have dramatic effects in very dense stellar cores.

At low and moderate energies, the fusion cross section of ``bare''
charged nuclei colliding with the relative momenta $p$ in the
center-of-mass frame is expressed as (see ref.~\cite{Lang}):
\begin{equation}
\label{sigma}
\sigma_b(E) = \frac{S(E)}{E}e^{-2\pi\eta},
\end{equation}
where $S(E)$ is the so-called astrophysical factor which incorporates all
nuclear features of the process, $E$ is the collision energy of the nuclei, 
$\eta = MZ_1Z_2/(m_e a_0 p)$ is the usual Coulomb parameter, $m_e$ and $M$
are the electron and reduced nuclear masses, respectively, and $a_0$ is the
hydrogen Bohr radius. The exponential factor originates from 
the Coulomb wave function of the internuclear motion
$\psi_E^{\mathrm C}(R)$ at $R = 0$.

The screened cross section $\sigma_s(E)$ differs by the enhancement factor
\begin{equation}
\label{gamma}
\gamma(E)\equiv \frac{\sigma_s}{\sigma_b} = 
\frac{|\psi_E(0)|^2}{|\psi_E^{\mathrm C}(0)|^2},
\end{equation}
where $\psi_E(R)$ is the wave function of the internuclear motion 
which accounts for the bound electron.

We evaluate the effect of electron screening of $^7$Be nucleus by one bound 
electron in reaction (\ref{creation}) in the framework of the 
adiabatic approximation for three particle problem. This calculation is
compared with two relevant approximations, ``united atom'' (UA) and
folding approximations which, as we will demonstrate below,
give respectively upper and lower estimates for the screening effect.
In the framework of the UA approximation we estimate also the 
screening effect for $^7$Be nucleus with two K-shell electrons.

\section{Method of Calculation}
\label{method}

We treat the Coulomb problem for three particles in the framework of the
adiabatic representation, that consists as follows. The three particle 
Schr\"odinger equation solution $\Psi({\mathbf r},{\mathbf R})$ is
represented as
\[
\Psi({\mathbf r},{\mathbf R}) = \sum_{k = 0}^\infty \psi_k({\mathbf R})
\phi_k({\mathbf r};{\mathbf R}).
\]
Here, ${\mathbf R}$ is the internuclear radius-vector, and ${\mathbf r}$ is
the electron radius-vector from the center-of-mass of nuclei.
The two-center eigenfunctions $\phi_k({\mathbf r};{\mathbf R})$ are derived
from the Schr\"odinger equation for three particles, when two nuclei are
fixed on a distance $R$, and the eigenfunctions have a dependence on $R$ as
a parameter.
The electron energy eigenvalue $U_{nlm}(R)$ also depends on the internuclear
distance.
The nuclei with electric charges $Z_1$ and $Z_2$ interact with the effective
potential
\begin{equation}
\label{potential}
V_{nlm}(R) = \frac{Z_1 Z_2}{R} + U_{nlm}(R),
\end{equation}
and $\psi_k({\mathbf R})$ are the eigenfunctions of nuclear motion in the 
potential 
(\ref{potential}). We shall write the parameters in atomic units, except that
we shall display the units. Thus, the energy unit is $ m_e e^4/\hbar^2 \approx
27.21$ eV, the lenght unit is $ a_0$, and the electric charge unit is $e$.

\begin{figure}
\center{\mbox{\epsfig{file=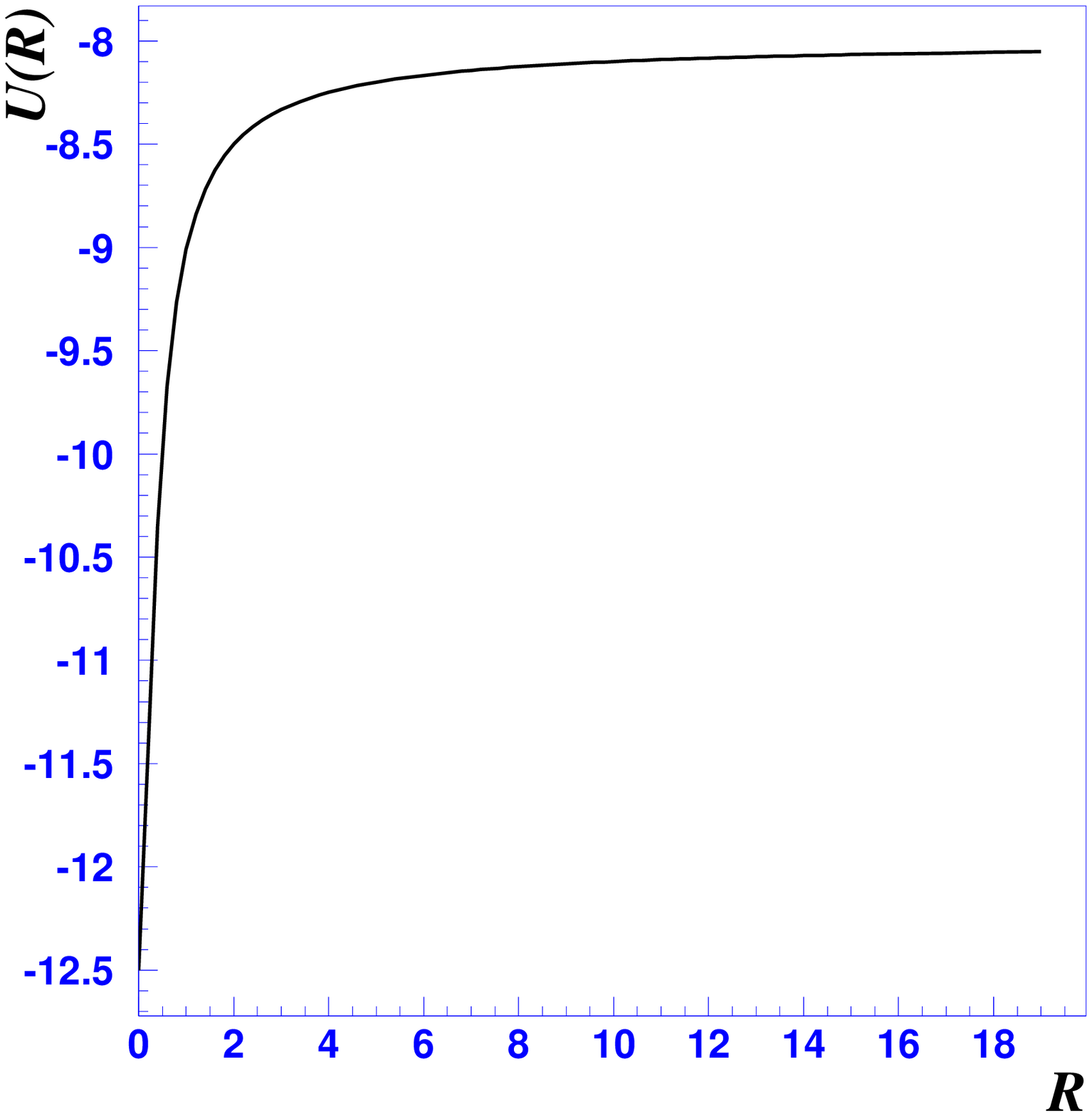,height=10cm}}}
        \protect\caption{Electron energy in the field of the two nuclei
                         $^7Be$ and $p$}
                         
\label{Beterm}
\end{figure}

Our approximation consists in using only one two-center eigenfunction
corresponding to the ground state of the system. Obviously, the main screening
effect is reached at zero orbital moment of the colliding nuclei. 
Thus, the fusing 
nuclei are considered in $S$-wave. 
There are some arguments for this approximation. 
At first, the high energy states corrections (at fixed $R$) are of the order 
of $\sqrt{m_e/M}$.
Then, excited energy levels correspond to the less energy of the united 
nucleus.
It leads to their exponentially small contribution into the nuclear fusion
rate in comparison with the ground state of the system.
At least, only the ground state energy of the electron in the field of two 
nuclei ($^7$Be and p in our case), labeled as $1S\sigma$ therm 
(see, for example ref.~\cite{Gershtein}) has the correct 
asymptotic behaviour (see below).
Thus, the total three particle wave function reads:
\[
\Psi({\mathbf r},{\mathbf R}) = \psi({\mathbf R}) 
\phi({\mathbf r};{\mathbf R}).
\]
The tabulated values of $U_{nlm}(R)$~\cite{Ponomarev} are used for
our purposes. The ground state eigenvalue energy $U(R)$ ($\equiv~U_{000}(R))$,
with zero quantum numbers, has the correct asymptotic behaviour: 
the electron eigenvalue energy reaches the energy of the 
united ion as $R$ approaches zero: $U(0) = (Z_1 + Z_2)^2/2$,
and $U(\infty) = Z_1^2/2$ (that is the energy of the 
isolated ion $eZ_1$).
In fig.~\ref{Beterm} the values for the ground state are plotted. 

We solved numerically the Schr\"odinger equation for two scattering nuclei
with $Z_1 =4$ and $Z_2 = 1$ at center-of-mass kinetic energy $E$ within the 
potential (\ref{potential}) with zero quantum numbers $n=l=m=0$.
The numerical solution was obtained on the mesh on $R$ for $R=0$ and
$R=R_{\max}$ in Numerov's scheme. 
We varied integration step $h$ and $R_{\max}$ in order to ensure that 
final result is not changed substantially. 
Also for checking purposes we reproduced the value of 
$|\psi_{E}^{\mathrm C}(0)|^2$ substituting $U(R) = 0$.
The values of the squared wave function of relative motion of the two nuclei
$|\psi_E(0)|^2$ at $0.1$ keV $\le E \le 100$ keV are calculated and compared
with $|\psi_{E}^{C}(0)|^2 = 2\pi\eta/(e^{2\pi\eta} - 1)$.

The UA approximation can serve as an upper limit to the screening
corrections induced by bound electron. The essence of this approximation
is replacement of the kinetic energy $E$ by the shifted energy 
$E + \Delta E$,
where $\Delta E$ is the energy difference of the bound electron between the
final and initial states. 
For one bound electron this energy difference is given by 
$\Delta E = (Z_1 + Z_2)^2/2 - Z_1^2/2$.
 
Thus, the UA approximation changes $|\psi_{E}^{\mathrm C}(0)|^2$
to $|\psi_{E + \Delta E}^{\mathrm C}(0)|^2$. 
Since the classical turning point for the ionic barrier penetration is at much
smaller radius than the Bohr radius ($a = a_0/Z_1$)  at solar energies,
the nuclear fusion takes place, in the main, within the united ion state.
The classical turning point is given by: 
\[
R_b = \frac{Z_1Z_2}{E}  \approx 
       a\left(\frac{0.4\;\mathrm{keV}}{E}\right),
\]
therefore, at solar energies, the
UA approximation applicability is well satisfied. Moreover, at
kinetic energies above few keV, $U(R)$ is not changed substantially at
$0 \le  R \le R_b$, and the UA approximation is expected to be
very close to the exact solution.
The UA approximation replaces decreasing $U(R)$ 
(see in fig.~\ref{Beterm}) by the constant  $\Delta E$, hence, this
approximation serves as an upper estimate.

The folding approximation can be used as a lower limit to the screening
corrections. This approximation embodies the static wave function of the bound
electron, which does not depend on the internuclear distance.
The total three particle wave function in this approximation is presented as 
a product:
\[
\Psi_f({\mathbf r},{\mathbf R}) = 
\psi_f({\mathbf R})\phi_f({\mathbf r}),
\]
where the wave function for the electron bounded on the nucleus with electric
charge $Z_1$ reads:
$
\phi_f({\mathbf r})=\sqrt{Z_1^3/\pi}e^{-Z_1r},
$
and $\psi_f({\mathbf R})$ is the wave function of the two nuclei.
Averaging  the electron coordinates, $\psi_f({\mathbf R})$ can be derived from
the  Schr\"odinger equation with the potential:
\begin{equation}
\label{static}
V_{f}(R) = 
\frac{Z_1Z_2}{R} - Z_2 \cdot 
\left(\frac{1-e^{-2RZ_1}}{R} - Z_1e^{-2RZ_1}\right).
\end{equation}
The zero point of the colliding nuclei kinetic energy is the electron 
eigenvalue energy in the field of $Z_1$ nucleus. 
Obviously, a rigid electron wave 
function, which has no dependence on $R$, can be used only in the case of
instantaneous impact of the nuclei. Thus, the area of the folding
approximation applicability is the faster velocity of the nuclei relative to
the electron one. 
Attractive part of the potential $V_{f}$ reaches the value: 
$Z_1Z_2$ as $R \longrightarrow 0$. 
Since $Z_1Z_2 \le (Z_1+Z_2)^2/2$ that is the value
of the electron energy in united ion, the folding approximation
gives a lower estimate for the enhancement factor. 

\section{Summary and discussion}
\begin{figure}
\center{\mbox{\epsfig{file=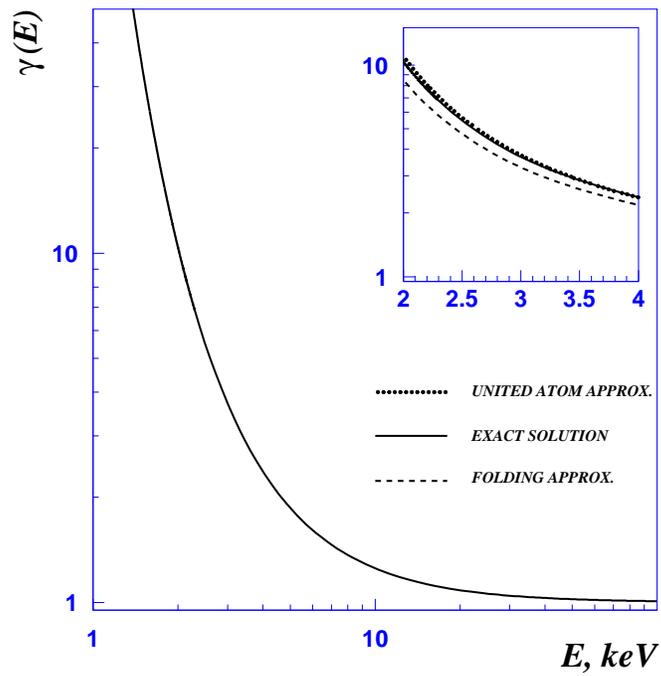,height=10cm}}}
        \protect\caption{Enhancement of nuclear fusion rate due to the
                         electron screening}
\label{enhan}
\end{figure}

The enhancement factor (\ref{gamma}) is plotted in fig.~\ref{enhan} for
all three approximations. As it was expected, the UA approximation
always overestimate the exact solution, while the folding approximation 
underestimates it. Nevertheless, it is easy to see that simple
UA prescription gives very close values to the exact solution at
kinetic energies above $2$ keV. Therefore, the latter can be used  not 
only as a qualitative, but as a good quantitative approximation to the electron
screening by bound electrons.

The electron screening is dramatic at very low kinetic energies of the nuclei.
However, in a plasma, most of the nuclear fusions come at the Gamow peak energy,
that is defined by the strong dependence of the nuclear cross section on energy
(\ref{sigma}) and the fast decrease of the exponential particle distribution.
This energy is given by:
\[
E_0 = 1.22(Z_1Z_2T_6)^{2/3}(M/M_p)^{1/3}\;{\rm keV},
\]
where $T_6 = T/10^6$ K, $M_p$ is the proton mass.
In the solar interior at $r_{\mathrm eff}/R_\odot = 0.06$, where  
the $^7$Be and $^8$B neutrino production reaches its maximum~\cite{BahPin},
the plasma parameters are $T_6 \approx 14.7$, the electron density 
$n_e \approx 7.7/a_{0}^3$, and the Gamow peak energy in reaction
(\ref{creation}) is about $18$ keV. Then $\gamma(E_0) = 1.1$, that is, there
is 10\% of an enhancement by one bound electron in the bohron production rate. 
Simple computations within the UA approximation give the
screening effect as:
\begin{equation}
\label{screen}
\gamma(E_0) = e^{\Delta E/kT}.
\end{equation}

Then, 10\% of an enhancement by one bound electron could be easily reproduced 
just inserting numbers into the formula (\ref{screen}).
One can apply this formula also for a $^7$Be nucleus with two bound electrons. 
Then, 
\[
\Delta E = \left(\chi_1^{\mathrm{B }}+\chi_2^{\mathrm{B }}\right) - 
              \left(\chi_1^{\mathrm{Be}}+\chi_2^{\mathrm{Be}}\right) =
               227.98 \: \mathrm{eV}.
\]
Here, $\chi_1^{\mathrm{B}} = 340.2$~eV and $\chi_2^{\mathrm{B}} = 259.4$~eV
are, respectively, the fifth and forth ionization potential of the $^8$B
atom and $\chi_1^{\mathrm{Be}} = 217.72$~eV and 
$\chi_2^{\mathrm{Be}} = 159.9$~eV are, respectively, the forth and third
ionization potential of the $^7$Be atom. 
Thus, two bound electrons enhancement factor is given by
$\gamma(E_0) = 1.196$, i.e. roughly 20\%.

Using the Saha equation Iben, Kalata and Schwartz~\cite{Iben} calculated the
probabilities $f_1$ and $f_2$ that one or 
two K-shell electrons are associated with any given $^7$Be nucleus. 
The calculations were perfomed under the assumption of pure Coulomb 
electron-ion forces, 
neglecting all excited states and screening. The probabilites found are
\begin{eqnarray*}
f_1 &=& \lambda\left[1+\lambda+0.25\lambda^2
        \exp{\left(-\frac{\Delta_\chi}{kT}\right)}\right]^{-1}, \\
f_2 &=& 0.25\lambda\exp{\left(-\frac{\Delta_\chi}{kT}\right)}f_1,
\end{eqnarray*}
where
\[
\lambda = n_e\left(\frac{h^2}{2\pi m_e kT}\right)^{3/2}
             \exp{\left(\frac{\chi_1^{\mathrm Be}}{kT}\right)}.
\]
Here $k$ is the  Boltzmann's constant, 
and $\Delta_{\chi}=\chi_1^{\mathrm{Be}}-\chi_2^{\mathrm{Be}}=63.8$~eV.
Inserting numbers one can obtain: $f_1 = 30\%$, and $f_2 = 3\%$.

Using the calculated abundances of $^7$Be ions, one can estimate
the thermal averaged screening effect induced by both one and two bound
electrons on a $^7$Be nucleus:
\[
\langle\gamma\rangle-1 = 0.30 \times 0.1 + 0.03 \times 0.2 \approx 0.04.
\]
In the Standard Solar  Model (SSM) the electron capture rate by $^7$Be
nucleus is taken to be about 1000 times faster than the proton capture 
rate~\cite{BahPin}. 
Therefore, a small change in $^8$B production rate does not affect 
significantly the $^7$Be neutrino flux, although it makes a proportional 
change in $^8$B neutrino flux. Thus, the electron screening by bound
electrons has the prompt consequences on bohron neutrino flux. 

The electron screening by plasma electrons from the continuum spectrum is 
expected
to contribute significantly to the total enhancement factor, since it is
proportional to $n_e$, and thus it has to be taken into account in the
exact prediction of neutrino flux change.

In summary, in the solar interior K-shell bound electrons enhance 
$^7$Be(p,$\gamma$)$^8$B rate
and increase $^8$B neutrino production rate by of about 4\%.
Therefore, bound electron screening has an effect on the solar $^8$B
neutrinos, and acts with the opposite effect to the berrylium neutrinos. 
The main essence of the electron screening in nuclear fusions is the change in
electron density on a nucleus during the collision of the nuclei. This effect
can be treated only in a dynamical calculation like the present three particle
calculation or the discussed UA approximation.  

We thank  J.~N.~Bahcall, A.~V.~Gruzinov and V.~A.~Naumov for usefull
discussions.


\newpage

\end{document}